\documentclass{desyproc}
\usepackage{here}

\newcommand{\alp}{\ensuremath{\alpha}}
\newcommand{\ardm}{ArDM}
\newcommand{\mgcs}{\ensuremath{\,\mathrm{mg/cm^{2}}}}

\begin{document}
\title{The Argon Dark Matter Experiment (\ardm )}

\author{{\slshape Christian Regenfus$^1$} on behalf of the \ardm\ collaboration\\[1ex]
$^1$Physik-Institut der Universit¬\"at Z¬\"urich, CH--8057 Z¬\"urich, Switzerland}  

\contribID{regenfus\_christian}

\desyproc{DESY-PROC-2008-02}
\acronym{Patras 2008} 
\doi  

\maketitle

\begin{abstract}

The \ardm\ experiment, a 1\,ton liquid argon TPC/Calorimeter,
is designed for the detection of dark matter particles which 
can scatter off the spinless argon nuclei. These events producing 
a recoiling nucleus will be discerned by their light to charge ratio, 
as well as the time structure of the scintillation light. The experiment 
is presently under construction and will be commissioned on 
surface at CERN. Here we describe the detector concept and 
give a short review on the main detector components.
\end{abstract}

\section{Introduction}
\label{sect:intro}
\vspace{-1mm}

Recent developments in 
noble liquid detectors give a promising outlook for this 
scalable technology which is favoured by high scintillation and 
ionisation yields. For the first time,
a liquid xenon detector\,\cite{xenon10} is producing limits that 
are compe\-ti\-tive with the currently well established dark 
matter searches of cryogenic semiconductor detectors\,\cite{cdms}. 
Gross target masses of around 50\,kg are currently employed and are 
pushing the effective exposures for experiments with event by event  
background recognition into the range of 100\,kg$\cdot$d. 
Upcoming larger noble liquid experiments\,\cite{xe100,warp,rubbia} 
will naturally have to fight more and more background events, not only for
the feedthrough in the phase space of the data but also in terms of maximal
tolerable trigger rates. However, selfshielding should improve 
performance for larger
\begin{wrapfigure}{r}{0.4\textwidth}
\vspace{-5mm}
\centerline{\includegraphics[width=0.4\textwidth]{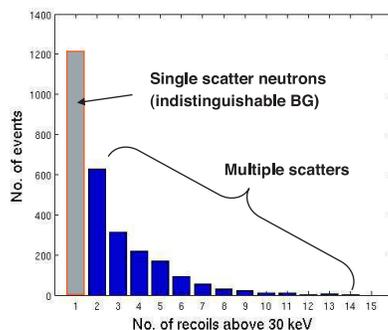}}
\vspace{-3mm}
\caption{Interaction multiplicity of background neutrons 
in \ardm\ (MC).}
\label{fig:nspect}
\vspace{-2mm}
\end{wrapfigure}
and larger target sizes, which is one of the strongest motivations to go to 
large masses. Above all, the number of single scattering neutrons
can then be determined (on a statistical base) from the distribution 
of multiple interacting neutrons of a given background spectrum.
Figure\,\ref{fig:nspect} shows the frequency distribution of 
neutron interactions on the example of the \ardm\ 
geometry\,\cite{lili}.

The best recognition of elec\-tron recoils (usually background) 
should generally be achieved in a two phase configuration of a noble 
liquid detector allowing for the multiplication and hence the
measurement of small ionisation charges.
In the case of liquid argon, both, the scintillation light to charge ratio
and the temporal structure of the light emission itself\,\cite{kubot,suzuk,hitach}
can be used for electron recoil discrimination. 
This is due to the ionisation-density-dependent po\-pu\-la\-tion of the two ground states
of argon excimers ($^{1}\Sigma^{+}_{u}$ and $^{3}\Sigma_{u}^{+}$) 
which are responsible for the VUV luminescense 
of li\-quid argon. The large ratio of their radiative lifetimes 
($\approx$\,10$^{2}$) allows for a good discrimination between electron 
and nuclear recoils, even at energies below 20\,keV on the electron equivalent 
scale\,\cite{macK}.  
A drawback of the argon technology is the short wavelength of 
the scintillation light (128\,nm) and the presence of 
the $^{39}$Ar $\beta$-emitter.
However, because of form factors, argon is less sensitive 
to the threshold of the nuclear recoil 
energy, than is e.g.\,xenon. For the same reason the 
recoil energy spectra of argon and xenon are quite different. 
These liquids are therefore complementary in providing a 
crosscheck once a WIMP signal has been found.

\section{Conceptual design}
\label{sect:design}

The main design goal of the \ardm\ project\,\cite{rubbia} was the construction
of a ton scale liquid argon detector for spectroscopy of nuclear 
recoils above 30\,keV. Three dimensional imaging and event by event 
interaction type identification will be used to reach a very 
high background suppression. An estimate of the final sensitivity
and its extrapolation to larger LAr projects is one of
the main subjects of this R\&D program which is a 
prototype unit for future large LAr detectors.
With current MC calculations we expect a sensitivity in the 
range of 10$^{-44}$\,cm$^{2}$ for a measurement of the spin 
independent cross section for weakly interacting dark 
matter particles.

In liquid argon about 400 VUV photons and a few free elementary 
charges\footnote{if the electrical field is above 1\,kV/cm}
are typically produced in a WIMP interaction at 30\,keV.
Background rejection will be achieved by the combination of
cuts on the fiducial volume, the event topology (e.g. no multiple 
scatter), the scintillation light to charge ratio, 
and the temporal structure of the light emission.
This requires a large homogeneous electric field over the full detector volume, 
a large area position sensitive charge readout (3rd dimension from drift time), 
a large area light readout with good time 
\begin{wrapfigure}{r}{0.45\textwidth}
\vspace{-3mm}
\centerline{\includegraphics[width=0.4\textwidth]{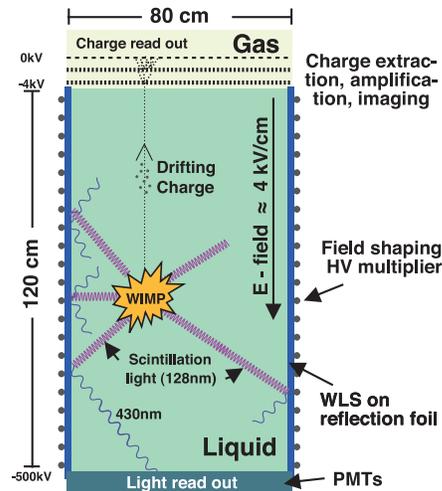}}
\vspace{-2mm}
\caption{Conceptual design of \ardm .}
\label{fig:concept}
\vspace{-2mm}
\end{wrapfigure}
resolution, and an efficient 
liquid argon purification 
system. The event trigger is generated by 
the fast light signal. Figure\,\ref{fig:concept} shows 
a sketch of the two-phase operating mode of the 
detector. An interaction in the liquid produces VUV
radiation (128\,nm) by a complicated process 
of excitation and ionisation of argon atoms, 
which results in the formation and subsequent radiative decay 
of the argon excimers\,\cite{suzuk}. This light 
can not be absorbed by neutral argon atoms and hence propagates to the side 
walls of the experiment which are coated with the wave shifting material 
tetra\-phenyl\-buta\-diene (TPB). The VUV light is absorbed 
and re-emitted with high efficiency at wavelengths around 430\,nm, 
which is a region of high quantum 
efficiency of standard bialkali photomultipliers (PMTs).
By diffusive reflection on the side walls, the light is transported
to the bottom of the apparatus where an array of 14 hemispherical 8'' 
PMTs is located. The strong electric field is capable of preventing 
free electrons in the densely ionised region around 
a nuclear recoil from recombining and sweeps them to the  surface 
of the liquid. Here there they are extracted into the gaseous phase 
and multiplied ($\approx$10$^{4}$) in the high
field regions ($\approx$30\,kV/cm) of a rigid large gas electron multiplier 
(LEM) which extends over the full detector surface. 

\section{Main experimental components and status}
\vspace{3mm}

\begin{figure}[h]
\centerline{
\includegraphics[width=0.42\textwidth]{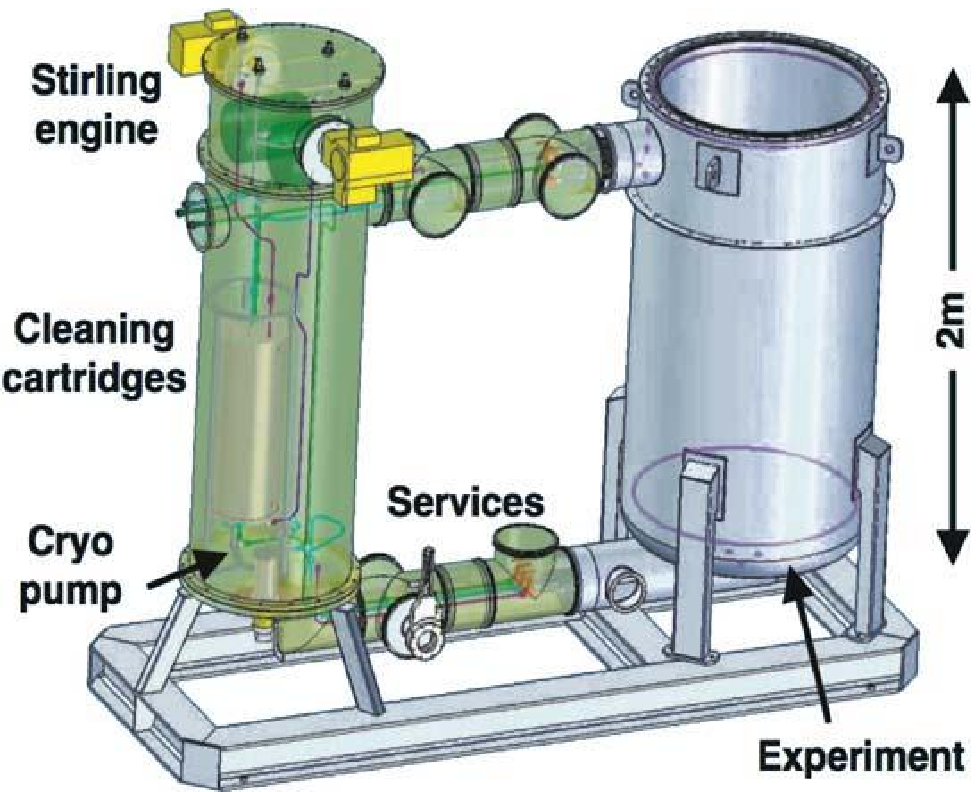}
\hspace{0.15\textwidth}
\includegraphics[width=0.23\textwidth]{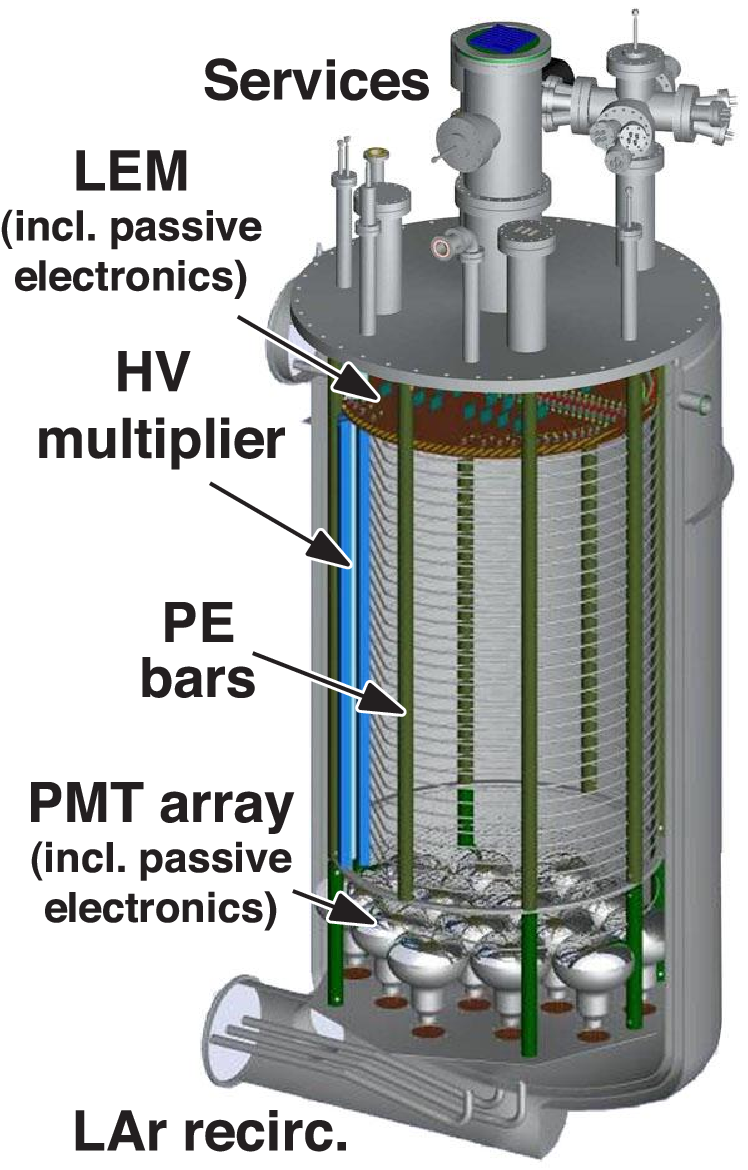}}
\vspace{4mm}
\caption{Left: 3D drawing of cryo-system and main dewar, 
right: view inside the main dewar with arrangement of the detector 
components hanging from the top flange on polyethylene bars.}
\label{fig:expdesign}
\end{figure}

Figure\,\ref{fig:expdesign} left shows the mechanical arrangement of the 
cryogenic cooling and cleaning system together with the main stainless steel 
dewar (containing roughly 1800\,kg of liquid argon). 
An inner cylindrical volume of 80\,cm diameter and 120\,cm height 
is delimited by round ring electrodes (field shapers). 
It is instrumented and 
used as a 850\,kg active LAr target in a vertical TPC
configuration (Fig.\,\ref{fig:expdesign} right). 
The field shaper rings are connected to a HV diode-capacitor 
charge pump system (Cockroft-Walton circuit) which is fully immersed in the liquid argon.
It consists of 210 stages and is designed to reach a voltage of -500\,kV at
its end creating an approximately 
\vspace{2mm}
4\,kV/cm vertical electric field. This design avoids an electrical HV 
feedthrough into the liquid phase, as well as a (lossy) voltage divider. 
The system  was 
tested successfully and takes advantage of the high dielectric 
breakdown voltage of liquid argon.

Another unique feature of the experiment is the realisation of
charge readout by a two-stage LEM manu\-factured by standard 
printed circuit board (PCB) methods. It will be placed 
5\,mm millimetres above the liquid level in the argon gas.
It consists of two 1.6\,mm thick 3\,mm spaced  Vetronite boards 
with holes of 0.5\,mm diameter and a readout anode. A stable 
overall gain of 10$^{3}$ is routinely attained by prototypes. 
The positional readout is achieved by segmenting the upper LEM 
surface and the anode plate with 1.5\,mm wide $x$ and $y$-strips 
respectively. In total there are 1024 readout channels 
which are AC coupled to charge sensitive preamplifiers 
located externally on the top flange of the apparatus. 
Because it is operated in pure argon gas, which cannot 
quench charge avalanches, the LEM is built with 
considerable attention to HV discharges.

\begin{wrapfigure}{l}{0.42\textwidth}
\vspace{-1mm}
\centerline{\includegraphics[width=0.36\textwidth]{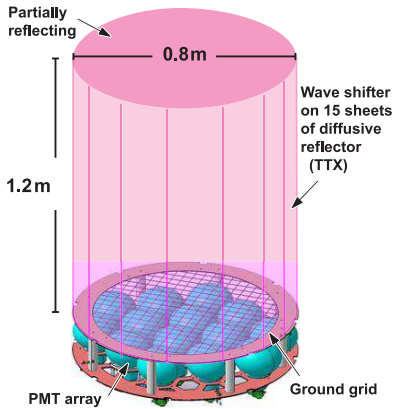}}
\vspace{-1mm}
\caption{\ardm\ light diffusion cell.}
\label{fig:lightro}
\vspace{-3mm}
\end{wrapfigure}

The readout for the 128\,nm scintillation light 
was designed with the constraint that large 
VUV sensitive photosensors (e.g.\,MgF$_2$ windowed PMTs) 
are commercially not available. 
To keep the system simple and scalable, we
chose an approach similar to a light diffusion cell
with an array of PMTs in the liquid argon at 
the bottom (Figure\,\ref{fig:lightro}).
To shift the light into the range of high quantum efficiency
of the borosilicate windowed bialkali PMTs, we evaporated 
a thin layer ($\approx$1\mgcs) of tetra\-phenyl\-buta\-diene 
(TPB) onto the 15, cylindrically arranged, 
25\,cm wide reflector 
sheets which are located in the vertical electric field. 
These sheets, which are made out of the PTFE fabric Tetratex{\small\texttrademark} 
(TTX), are clamped to the upper- and lowermost field 
shaper rings. The Hamamatsu PMTs R5912-02MOD-LRI, sensitive to single photons,
are made from particularly radiopure 
borosilicate glass and feature bialkali photocathodes with Pt-underlay 
for operation at cryogenic temperatures. This ineluctably reduces their 
quantum efficiency by roughly one third to a value of appro\-xi\-ma\-tely 
15\%\,\cite{granada}.
The PMT glass windows are also coated with TPB to convert directly 
impinging VUV photons. The average number of detected photoelectrons 
per produced 128\,nm photon of the overall detector is currently under
investigation. From laboratory measurements we expect a value around 
2-5\%. The development of this light detection system 
and particularly the operation of (gaseous) argon test cells 
with \alp\ particle excitation were described 
in earlier work\,\cite{vuvidm,lumquench}. A more detailed 
description of the present experimental state can be 
found in\,\cite{calor}. 

\vspace{-1mm}
\section*{Outlook}
\vspace{-1mm}

While R\&D work for sub detector parts is finalising, the main 
mechanical components are set together on surface at CERN, 
allowing for their commissioning. Following a successful 
operation at surface and later on at shallow depth (CERN), we 
consider a deep underground operation. 

\vspace{-2mm}
\section*{Acknowledgements}
\vspace{-1mm}

This work is supported by grants from the Swiss National 
Science Foundation and ETH Z¬\"urich.
 
\vspace{-1mm}
\begin{footnotesize}

\end{footnotesize}

\end{document}